\documentclass[12pt]{article}
\usepackage{epsf,pst-plot}
\newcommand\pubnumber{DESY 01-028}
\newcommand\pubdate{March 7, 2001 \\ December 9, 2002}
\newcommand\hepnumber{~}

\def\csumb{Deutsches Elektronen-Synchrotron DESY \\
Platanenallee 6, \\ D--15738 Zeuthen, Germany}
\def\support{\footnote{Presented at ``50 Years of Electroweak Physics''
a symposium in honor of Professor 
Alberto Sirlin's 70th Birthday at New York University, New York,
October 27-28, 2000\\
(enhanced by 3 figures and an addendum; updated Dec. 2002)\\
to appear in {\it J. Phys. G: Nucl. Part. Phys.} {\bf 28} (2002).  
}} 

\def\Title#1{\begin{center} {\Large\bf #1 } \end{center}}
\def\Author#1{\begin{center}{ \sc #1} \end{center}}
\def\Address#1{\begin{center}{ \it #1} \end{center}}

\newcommand\pubblock{\rightline{\begin{tabular}{l} \pubnumber\\
         \pubdate\\ \hepnumber \end{tabular}}}
\newenvironment{Abstract}{\begin{quotation}  }{\end{quotation}}

\makeatletter
\def\section{\@startsection{section}{0}{\z@}{5.5ex plus .5ex minus
 1.5ex}{2.3ex plus .2ex}{\large\bf}}
\def\subsection{\@startsection{subsection}{1}{\z@}{3.5ex plus .5ex minus
 1.5ex}{1.3ex plus .2ex}{\normalsize\bf}}
\def\subsubsection{\@startsection{subsubsection}{2}{\z@}{-3.5ex plus
-1ex minus  -.2ex}{2.3ex plus .2ex}{\normalsize\sl}}

\renewcommand{\@makecaption}[2]{%
   \vskip 10pt
   \setbox\@tempboxa\hbox{\small #1: #2}
   \ifdim \wd\@tempboxa >\hsize     
       \small #1: #2\par          
     \else                        
       \hbox to\hsize{\hfil\box\@tempboxa\hfil}
   \fi}

 \def\citenum#1{{\def\@cite##1##2{##1}\cite{#1}}}
\def\citea#1{\@cite{#1}{}}
 
\newcount\@tempcntc
\def\@citex[#1]#2{\if@filesw\immediate\write\@auxout{\string\citation{#2}}\fi
  \@tempcnta\z@\@tempcntb\m@ne\def\@citea{}\@cite{\@for\@citeb:=#2\do
    {\@ifundefined
       {b@\@citeb}{\@citeo\@tempcntb\m@ne\@citea\def\@citea{,}{\bf ?}\@warning
       {Citation `\@citeb' on page \thepage \space undefined}}%
    {\setbox\z@\hbox{\global\@tempcntc0\csname b@\@citeb\endcsname\relax}%
     \ifnum\@tempcntc=\z@ \@citeo\@tempcntb\m@ne
       \@citea\def\@citea{,}\hbox{\csname b@\@citeb\endcsname}%
     \else
      \advance\@tempcntb\@ne
      \ifnum\@tempcntb=\@tempcntc
      \else\advance\@tempcntb\m@ne\@citeo
      \@tempcnta\@tempcntc\@tempcntb\@tempcntc\fi\fi}}\@citeo}{#1}}
\def\@citeo{\ifnum\@tempcnta>\@tempcntb\else\@citea\def\@citea{,}%
  \ifnum\@tempcnta=\@tempcntb\the\@tempcnta\else
  {\advance\@tempcnta\@ne\ifnum\@tempcnta=\@tempcntb \else\def\@citea{--}\fi
    \advance\@tempcnta\m@ne\the\@tempcnta\@citea\the\@tempcntb}\fi\fi}
\makeatother

\def\NPB{{\em Nucl. Phys.} B }

\def\PLB{{\em Phys. Lett.}  B }

\def\PRD{{\em Phys. Rev.} D }
\def\ZPC{{\em Z. Phys.} C }

\def\amu{a_\mu}
\def\amuh{a_\mu^{{\mathrm had}}}
\def\MZ{M_Z}

\def\az{\alpha(\MZ)}

\def\dalf{\Delta\alpha}
\def\das{\Delta\alpha(s)}

\def\dah{\Delta\alpha^{(5)}_{\rm had}}

\def\dahz{\Delta\alpha^{(5)}_{\rm had}(\MZ^2)}
\def\dahzE{\Delta\alpha^{(5)}_{\rm had}(-\MZ^2)}
\def\dah0{\Delta\alpha^{(5)}_{\rm had}(-s_0)}

\def\damu{\delta \amu}

\newcommand{\gv}{\mbox{GeV}}

\newcommand{\MOM}{${\mathrm{MOM}}$ }

\newcommand{\MSb}{$\overline{\mathrm{MS}}$ }

\newcommand{\al }{\alpha}
\newcommand{\epm}{e^+e^-}
\newcommand{\pipi}{\pi^+\pi^-}

\newcommand{\be}{\begin{equation}}
\newcommand{\ee}{\end{equation}}
\newcommand{\ba}{\begin{eqnarray}}
\newcommand{\ea}{\end{eqnarray}}
\newcommand{\bea}{\begin{eqnarray*}}
\newcommand{\eea}{\end{eqnarray*}}
\newcommand{\bet}{\begin{center} \begin{tabular}}
\newcommand{\ent}{\end{tabular} \end{center}}
\newcommand{\bb}{\begin{thebibliography}}
\newcommand{\eb}{\end{thebibliography}}

\newcommand{\ra}{\rightarrow}

\newcommand{\bit}{\begin{itemize}}
\newcommand{\eit}{\end{itemize}}

\newcommand{\veps}{\varepsilon}

\newcommand{\lapprox}{\raisebox{-.2ex}{$\stackrel{\textstyle<}
{\raisebox{-.6ex}[0ex][0ex]{$\sim$}}$}}

\newcommand{\crn}{\nonumber \\}
\newcommand{\nn}{\nonumber}
\newcommand{\ha}{\frac{1}{2}}
\newcommand{\dal}{\Delta \alpha}

\newcommand{\sha}{\sigma(e^+e^- \rightarrow {\rm hadrons})}
\newcommand{\mz}{M^2_Z}

\newcommand{\de}{\frac{\delta e}{e}}

\newcommand{\sinf}{\sin^2 \Theta_f}
\newcommand{\cosf}{\cos^2 \Theta_f}
\newcommand{\sini}{\sin^2 \Theta_i}
\newcommand{\cosi}{\cos^2 \Theta_i}
\newcommand{\sinW}{\sin^2 \Theta_W}
\newcommand{\cosW}{\cos^2 \Theta_W}
\newcommand{\sing}{\sin^2 \Theta_g}

\newcommand{\dro}{\Delta \rho}
\newcommand{\Gmu}{G_{\mu}}
\newcommand{\bary}{\begin{array}}
\newcommand{\eary}{\end{array}}

\begin{document}
\begin{titlepage}
\pubblock

\vfill
\def\thefootnote{\fnsymbol{footnote}}
\Title{Hadronic Contributions to the
Photon Vacuum Polarization \\[5pt]
and their Role in Precision Physics}
\vfill
\Author{Fred Jegerlehner\support}
\Address{\csumb}
\vfill
\begin{Abstract}
I review recent evaluations of the hadronic contribution to the shift
in the fine structure constant and to the anomalous magnetic moment of
the muon. Substantial progress in a precise determination of these
important observables is a consequence of substantially improved total
cross section measurement by the CMD-2 and BES II collaborations and
an improved theoretical understanding. Prospects for further possible
progress is discussed.
\end{Abstract}
\vfill
\end{titlepage}
\def\thefootnote{\arabic{footnote}}
\setcounter{footnote}{0}

\section{Introduction}

Precision physics requires appropriate inclusion of higher order
effects and the knowledge of very precise input parameters of the
electroweak Standard Model SM. One of the basic input parameters is
the fine structure constant which depends logarithmically on the
energy scale. Vacuum polarization effects lead to a partial screening
of the charge in the low energy limit (Thomson limit) while at higher
energies the strength of the electromagnetic interaction grows.  We
discuss the current status of the hadronic contributions to some
electroweak precision observables like the leading hadronic
contribution to the muon anomalous magnetic moment $a_\mu \equiv
(g_\mu-2)/2$~\cite{BNL} and the effective fine structure constant at
the $Z$--resonance~\cite{LEP}.

Renormalization of the electric charge $e$ by a shift $\delta
e$ at different scales leads to a shift of the fine structure
constant by
\be
\Delta \alpha = 2 \left( \de (0) - \de (M_Z)\right)
= \Pi'_{\gamma}(0)-\Pi'_{\gamma}(\mz)
\ee
where $\Pi'_{\gamma}(s)$ is the photon vacuum polarization function
defined via the time-ordered product of two electromagnetic currents
$j_{em}^\mu(x)$:
\be
i\int d^4x\,e^{iq\cdot x}\langle 0| {\rm T} j_{em}^\mu(x) j_{em}^\nu(0)
                        |0 \rangle
= -(q^2 g^{\mu\nu} - q^\mu q^\nu)\Pi'_{\gamma}(q^2)\;.
\ee
The shift $\Delta \alpha$ is large due to the large change in scale
going from zero momentum to the Z-mass scale $\mu=M_Z$ and due to the
many species of fermions contributing. Zero momentum more precisely
means the light fermion mass thresholds.

In perturbation theory the leading light fermion ($m_f \ll M_Z$)
contribution is given by

\hfill \epsfbox{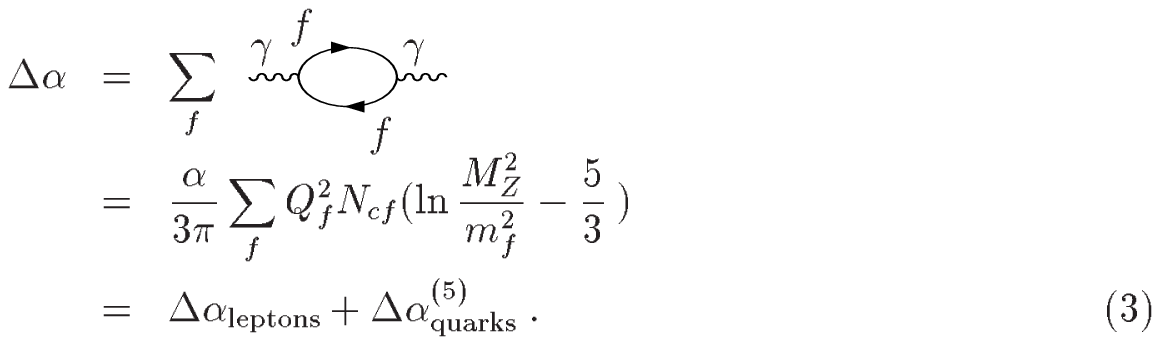}

\setcounter{equation}{3}

A serious problem is the low energy contributions of the five light
quarks u,d,s,c and b which cannot be reliably calculated using
perturbative quantum chromodynamics (p-QCD). The evaluation of the
hadronic contribution $\Delta \al _{\rm quarks}^{(5)} \ra \Delta \al
_{\rm hadrons}^{(5)}$ is the main concern of this mini review. Before I
am going into this, let me make a few remarks about the consequences of
the related problems for precision physics.

A major drawback of the partially non-perturbative relationship
between $\alpha(0)$ and $\alpha(M_Z)$ is that one has to rely on
experimental data exhibiting systematic and statistical errors which
implies a non-negligible uncertainty in our knowledge of the
effective fine structure constant. In precision predictions of gauge
boson properties this has become a limiting factor.  Since $\alpha~,~
G_\mu, M_Z$ are the most precisely measured parameters, they are used
as input parameters for accurate predictions of observables like the
effective weak mixing parameter $\sinf$, the vector $v_f$ and
axial-vector $a_f$ neutral current couplings, the $W$ mass $M_W$ the
widths $\Gamma_Z$ and $\Gamma_W$ of the $Z$ and the $W$, respectively,
etc. However, for physics at higher energies we have to use the
effective couplings at the appropriate scale, for physics at the
$Z$--resonance, for example, $\alpha(M_Z)$ is more adequate to use
than $\alpha(0)$. Of course this just means that part of the higher
order corrections may be absorbed into an effective parameter. If we
compare the precision of the basic parameters
\be \bary{ccccccccccc}
\frac{\delta \alpha}{\alpha} &\sim& 3.6 &\times& 10^{-9}&~~~~~&
\frac{\delta \alpha(M_Z)}{\alpha(M_Z)} &\sim& 1.6 \div 6.8 &\times& 10^{-4}\\
\frac{\delta G_\mu}{G_\mu} &\sim& 8.6 &\times& 10^{-6}&&
\frac{\delta M_Z}{M_Z} &\sim& 2.4 &\times& 10^{-5}
\eary
\ee
we observe that the uncertainty in $\alpha(M_Z)$ is roughly an order
of magnitude worse than the next best, which is the $Z$--mass. Let me
remind the reader that $\dal$ enters in electroweak precision physics
typically when calculating versions of the weak mixing parameter
$\sini$ from $\al$, $\Gmu$ and $M_Z$ via
\be
\sini\:\cosi\: =\frac{\pi \al}{\sqrt{2}\:G_\mu\:M_Z^2}
\frac{1}{1-\Delta r_i}\:.
\ee
where
\ba
\Delta r_i &=&\Delta r_i({\al ,\: \Gmu ,\: M_Z ,}\:m_H,
\:{m_{f\neq t},\:m_t})
\ea
includes the higher order corrections which can be calculated in the
SM or in alternative models. It has been calculated for the first time
by Alberto Sirlin in 1980~\cite{Sirlin80}. In the SM today the Higgs
mass $m_H$ is the only relevant unknown parameter and by confronting
the calculated with the experimentally determined value of $\sini$ one
obtains important indirect constraints on the Higgs mass. $\Delta
r_i$ depends on the definition of $\sini$. The various definitions
coincide at tree level and hence only differ by quantum effects. From
the weak gauge boson masses, the electroweak gauge couplings and the
neutral current couplings of the charged fermions we obtain
\ba
\sinW &=& 1-\frac{M_W^2}{M_Z^2}\\
\sing &=& e^2/g^2=\frac{\pi \al}{\sqrt{2}\:G_\mu\:M_W^2}\\
\sinf &=&
\frac{1}{4|Q_f|}\;\left(1-\frac{v_f}{a_f} \right)\;,\;\;f\neq \nu\;,
\ea
for the most important cases and the general form of $\Delta r_i$ reads
\ba
\Delta r_i &=& \dal - f_i(\sini)\:\dro + \Delta r_{i\:\mathrm{reminder}}
\ea
with a universal term $\dal$ which affects the predictions of
$M_W$, $A_{LR}$, $A^f_{FB}$, $\Gamma_f$, etc. The uncertainty $\delta
\Delta \alpha$ implies uncertainties $\delta M_W$, $\delta \sini$
given by
\ba
\frac{\delta M_W}{M_W} &\sim& \ha \frac{\sinW}{\cosW-\sinW}
\;\delta \dal \sim 0.23 \;\delta \dal \\
\frac{\delta \sinf}{\sinf} &\sim& ~~\frac{\cosf}{\cosf-\sinf}
\;\delta \dal \sim 1.54 \;\delta \dal\;
\ea


\vspace*{46mm}

\begin{figure}[h]
\begin{picture}(120,60)(-20,0)
{\scalebox{0.75 0.75}{%
\epsfbox{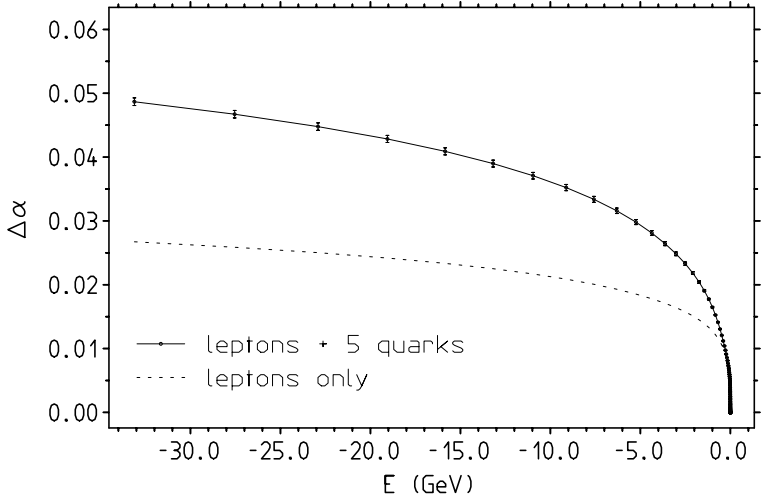}}}
\end{picture}

\caption[]{The running of $\alpha$. The ``negative'' $E$ axis is
chosen to indicate space-like momentum transfer. The vertical bars at
selected points indicate the uncertainty.}
\label{fig:alpharun}
\end{figure}
which obscure in particular the indirect bounds on the Higgs mass
obtained from electroweak precision measurements. A summary of the
present status and future expectations will be presented below. Once
the Higgs boson will have been discovered and its mass is known, precision
measurements of the $\Delta r_i$, which would be possible with the
GigaZ option of TESLA~\cite{TDR}, would provide excellent possibilities
to establish new physics contributions beyond the SM.

\section{The hadronic contributions to $\alpha(s)$}

The effective QED coupling constant at scale $\sqrt{s}$ is given by
the renormalization group resummed running fine structure constant
\be
\alpha(s) =  \frac{\alpha}{1 - \das}
\ee
with
\be
\das =- 4\pi\alpha
        {\rm Re}\left[ \Pi'_{\gamma}(s) - \Pi'_{\gamma}(0)
                \right]\;.
\ee
Fig.~\ref{fig:alpharun} illustrates the running of the effective
charges at lower energies in the space-like region.  Typical values
are $\Delta \al (5 \gv) \sim 3\%$ and $\Delta \al (M_Z)
\sim 6\%$, where about $\sim 50\%$ of the contribution comes from 
leptons and about $\sim 50\%$ from hadrons.

The leptonic contributions are calculable in perturbation theory where
at leading order the free lepton loops yield
\be \bary{l}
\dalf_{{\rm leptons}}(s)= \cr \bary{lcl}
& = & \sum\limits_{\ell=e,\mu,\tau}
      \frac{\alpha}{3\pi}
      \left[ - \frac{8}{3} + \beta_\ell^2
             - \frac{1}{2}\beta_\ell(3 - \beta_\ell^2)
               \ln\left( \frac{1-\beta_\ell}{1+\beta_\ell}
                  \right)
      \right]   \cr
& = &
      \sum\limits_{\ell=e,\mu,\tau}
      \frac{\alpha}{3\pi}
      \left[ \ln\left( s/m_\ell^2
                \right)
           - \frac{5}{3}
           + O\left( m_\ell^2/s
              \right)
      \right]  {\rm \ for \ } |s|\gg m_\ell^2   \cr
& \simeq & 0.03142 {\rm \ for \ } s=M_Z^2 \eary \eary
\ee
where $\beta_\ell = \sqrt{1 - 4m_\ell^2/s}$. This leading contribution
is affected by small electromagnetic corrections only in the next to
leading order. The leptonic contribution is actually known to three
loops~\cite{KalSab55,Ste98} at which it takes the value
\be
\Delta \alpha_{\rm leptons} (M_Z^2) \; \simeq \; 314.98 \: \times \: 10^{-4}.
\ee

In contrast, the corresponding free quark loop contribution gets
substantially modified by low energy strong interaction effects, which
cannot be obtained by p-QCD.  Fortunately, one can evaluate this
hadronic term $\Delta \al _{\rm hadrons}^{(5)}$ from hadronic $\epm $-
annihilation data by using a dispersion relation. The relevant vacuum
polarization amplitude satisfies the convergent dispersion relation
\bea
Re \Pi'_{\gamma}(s)-\Pi'_{\gamma}(0)=\frac{s}{\pi} Re \int_{s_0}^{\infty}
ds' \frac{Im \Pi'_{\gamma}(s')}{s'(s'-s -i \veps)} \nn
\eea
and using the optical theorem (unitarity) one has
\bea
Im \Pi'_{\gamma}(s)=\frac{s}{e^2}
                    \sigma_{tot} (\epm \ra \gamma^* \ra {\rm hadrons})(s)\;. \nn
\eea
In terms of the cross-section ratio
\be
R(s)=\frac{\sigma_{tot} (\epm \ra \gamma^* \ra {\rm hadrons})}
     {\sigma (\epm \ra \gamma^* \ra \mu^+ \mu^-)}\;, \nn
\label{RS}
\ee
where $\sigma (\epm \ra \gamma^* \ra \mu^+ \mu^-)=\frac{4\pi
\al ^2}{3s}$ at tree level, we finally obtain
\be
\Delta \al _{\rm hadrons}^{(5)}(M_Z^2) =
-\frac{\al M_Z^2}{3\pi}Re\int_{4m_{\pi}^2}^{\infty}
ds\frac{R(s)}{s(s-M_Z^2-i\veps)}\;.
\label{DA}
\ee
Using the experimental data for $R(s)$ up to $\sqrt{s}=E_{cut}=5$
GeV and for the $\Upsilon$ resonances region between 9.6 and 13 GeV
and perturbative QCD from 5.0 to 9.6 GeV and for the high energy
tail~\cite{GKL,ChK95,ChHK00} above 13 GeV we get as an update
of~\cite{EJ95} including the recent new data from CMD~\cite{CMD} and
BES~\cite{BES}
\bea
\Delta \al _{\rm hadrons}^{(5)}(\mz) &=& 0.027572 \pm 0.000359\;\;;\;\;\;
\alpha^{-1}(\mz)=128.952 \pm 0.049  
\eea
at $M_Z=$ 91.19 GeV (see also~\cite{FJ01}). The CMD-2 experiment at Novosibirsk
has continued and substantially improved to 0.6\% the $\sha$
measurements below 1.4 GeV~\cite{CMD} and the BES II experiment at
Beijing has published a new measurement, which in the region from 2 to
5 GeV improves the evaluation from 15\% to 20\% systematic error to
about 6.6\%~\cite{BES}. As a consequence we observe a dramatic
reduction of the error with respect to our 1995 evaluation $0.0280\pm
0.0007$~\cite{EJ95} mainly due to the new BES data (see also~\cite{BP01}) . 
A number of recent evaluations are summarized in Tab.~\ref{tab:alperr} below.

\section{Theoretical progress}

Because of the large uncertainties in the data many authors advocated
to extend the use of perturbative QCD in place of
data~\cite{DH98a,KS98,GKNS98,DH98b,Erler98,MOR00}. The assumption that p-QCD
may be reliable to calculate $R(s)$ down to energies as low as 1.8 GeV
seems to be supported by
\begin{itemize}
\item the apparent applicability of p-QCD to $\tau$ physics. In fact the
running of $\alpha_s(M_\tau) \ra \alpha_s(M_Z)$ from the $\tau$ mass
up to LEP energies agrees well with the LEP value. The estimated
uncertainty may be debated, however.
\item the smallness of
non--perturbative (NP) effects~\cite{DH98b} (see also:~\cite{FJ86}) if
parameterized as prescribed by the operator product expansion (OPE) of
the electromagnetic current correlator~\cite{SVZ}.
\end{itemize}
Progress in p-QCD comes mainly from~\cite{mqcd3}. In addition an exact
two--loop calculation of the renormalization group (RG) in the gauge
invariant background field \MOM scheme is now available~\cite{JT98}
which allows us to treat ``threshold effects'' closer to physics than
in the \MSb scheme. Except from Ref.~\cite{KS98} which is based
on~\cite{mqcd3} most other ``improved'' calculations utilize older
results, mainly, the well known massless result~\cite{GKL} plus some
leading mass corrections. For a recent critical review of the newer
estimates of vacuum polarization effects see~\cite{FJ98} and
Tab.\ref{tab:alperr} below.

\begin{figure}[h]
\begin{picture}(120,60)(-20,0)
\rput{00}(6.4,3.7){\scalebox{.9 .9}{%
\epsfbox{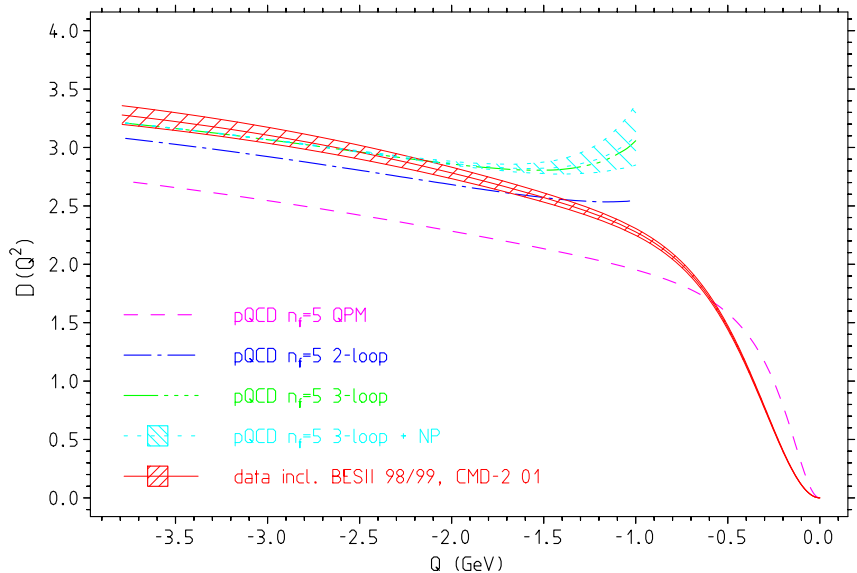}}}
\end{picture}

\caption[]{The Adler function: theory vs. experiment.}
\label{fig:Adlerb}
\end{figure}

In Ref.~\cite{EJKV98} a different approach of p-QCD improvement was
proposed, which relies on the fact that the vacuum polarization
amplitude $\Pi(q^2)$ is an analytic function in $q^2$ with a cut in
the $s$--channel $q^2=s \geq 0$ at $s\geq 4m^2_\pi$ and a smooth
behavior in the $t$--channel (space-like or Euclidean region). Thus,
instead of trying to calculate the complicated function $R(s)$, which
obviously exhibits non-perturbative features like resonances, one
considers the simpler Adler function in the Euclidean
region. In~\cite{EJKV98} the Adler function was investigated and p-QCD
was found to work very well above 2.5 GeV, provided the exact
three--loop mass dependence was used (in conjunction with the
background field \MOM scheme). The Adler function may be defined as a
derivative
\be
D(-s)=-(12\pi^2)\,s\,\frac{d\Pi'_{\gamma}\,(s)}{ds}
=\frac{3\pi}{\alpha} s\frac{d}{ds}\Delta \alpha_{\mathrm{had}}(s)
\label{DD}
\ee
of (\ref{RS}) which is the hadronic contribution to the shift of the
fine structure constant. It is represented by
\ba
D(Q^2)=Q^2\:\left(\int_{4 m_{\pi}^2}^{E^2_{\rm cut}}
\frac{  R^{\rm data}(s)}{(s+Q^2)^2}ds\;+\;
\int_{E^2_{\rm cut}}^{\infty}\frac{  R^{\rm pQCD}(s)}{(s+Q^2)^2}ds\:\right)
\label{DI}
\ea
in terms of the experimental $\epm$--data. The standard evaluation
(\cite{EJ95}) of (\ref{DI}) then yields the non--perturbative
``experimental'' Adler function, as displayed in
Fig.~\ref{fig:Adlerb}.

For the p-QCD evaluation it is mandatory to utilize the calculations
with massive quarks which are available up to
three--loops~\cite{mqcd3}.  The four-loop corrections are known in the
approximation of massless quarks~\cite{GKL}. The outcome of this
analysis is pretty surprising and is shown in Fig.~\ref{fig:Adlerb}.
For a discussion we refer to the original paper~\cite{EJKV98}.


\vspace*{56mm}

According to (\ref{DD}), we may compute the hadronic vacuum
polarization contribution to the shift in the fine structure constant
by integrating the Adler function. In the region where p-QCD works
fine we integrate the p-QCD prediction, in place of the data. We thus
calculate in the Euclidean region
\be
\Delta\alpha^{(5)}_{\rm had}(-M_Z^2)
=\left[\Delta\alpha^{(5)}_{\rm had}(-M_Z^2) -\Delta\alpha^{(5)}_{\rm
had}(-s_0)\right]^{\mathrm{p-QCD}}+ \Delta\alpha^{(5)}_{\rm
had}(-s_0)^{\mathrm{data}}\;\;.
\ee
A save choice is $s_0=(2.5\, \gv)^2$ where we obtain
$\Delta\alpha^{(5)}_{\rm had}(-s_0)^{\mathrm{data}} =0.007340 \pm
0.000093$ from the evaluation of the dispersion integral (\ref{RS}).
With the results presented above we find~\cite{FJ01}
\ba
\Delta\alpha^{(5)}_{\rm
had}(-M_Z^2) = 0.02745 \pm 0.00018 
\ea
for the Euclidean effective fine structure constant. Adding $\Delta\alpha^{(5)}_{\rm
had}(M_Z^2)-\Delta\alpha^{(5)}_{\rm
had}(-M_Z^2)$ we obtain
\ba
\Delta\alpha^{(5)}_{\rm
had}(M_Z^2) = 0.02737 \pm 0.00020\;\;.
\ea
In~\cite{FJ01} I have evaluated the uncertainty of the (in this
approach) large p-QCD part. Uncertainties in the strong coupling
constant and in the quark masses are equally important and yield a
substantial error $\delta \Delta\alpha^{(5)}_{\rm had}(-M_Z^2) =
0.00015$. Table~\ref{tab:alperr} compares our results with results
obtained by other authors which obtain smaller errors because they are
using p-QCD in a less conservative manner.

\begin{table}[ht]
\begin{tabular}{ll|c|c|cc}
\hline
$\dahz$ &$\delta \dal$ & $\delta \sinf$& $\delta M_W$ & Method & Ref. \\
\hline
0.0280~&0.00065~~ & 0.000232 & 12.0 & data $< ~12.~~\gv$                      &\cite{EJ95} \\
0.02777&0.00017~~ & 0.000061 &  3.2 & data $< ~1.8~~\gv$                      &\cite{KS98}\\
0.02763&0.00016~~ & 0.000057 &  3.0 & data $< ~1.8~~\gv$                      &\cite{DH98b}\\
0.027572 & 0.000359  & 0.000128 &  6.6 &  ~\cite{EJ95} + new data
CMD \& BES      &\cite{FJ01,FJ02} \\
0.02737&0.00020 & 0.000071 &  3.7 & Euclidean $> ~2.5~~\gv$                 &\cite{FJ98} \\
0.027426&0.000190  & 0.000070 &  3.6 & scaled data, pQCD 2.8-3.7, 5-$\infty$  &\cite{MOR00}\\
0.027649&0.000214  & 0.000078 &  4.0 & same but ``exclusive''                  &\cite{MOR00}\\
0.02761&0.00036  & 0.000128 &  6.6 & data-driven incl. BES    &\cite{BP01}\\
~~~~~-&0.00007~~ & 0.000025 &  1.3 &{$\delta \sigma \:\lapprox\: 1\%$ up to $J/\psi$}\\
~~~~~-&0.00005~~ & 0.000018 &  0.9 &{$\delta \sigma \:\lapprox\: 1\%$ up to $\Upsilon$}\\
\multicolumn{2}{c|}{world average} & 0.00017 & 39.0 & PDG 2002 \\
\hline
\end{tabular}
\caption{%
$\dahz$ and its uncertainties in different evaluations. Two
entries show what can be reached by increasing the precision of cross
section measurements to 1\%. $\delta M_W$ in MeV.}
\label{tab:alperr}
\end{table}

\begin{figure}[hb]
\begin{picture}(120,60)(30,20)
\epsfbox{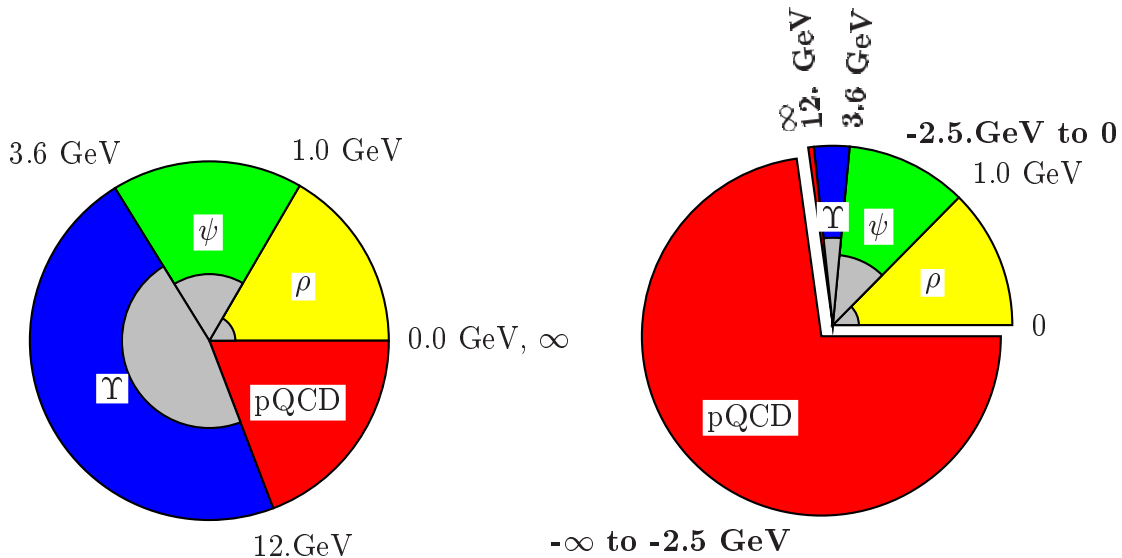}
\end{picture}

\caption{Comparison of the distribution of contributions and errors
(shaded areas scaled up by 10) in the standard (left) and the Adler
function based approach (right), respectively.}
\label{fig:alpsta}
\end{figure}

Our procedure to evaluate $\dahzE$ in the Euclidean region has
several advantages as compared to other approaches used so far:
The virtues of our analysis are the following:
\begin{itemize}
\item no problems with the physical threshold and resonances
\item p-QCD is used only in the Euclidean region and not below 2.5 GeV.
 For lower scales p-QCD ceases to describe properly the functional
 dependence of the Adler function~\cite{EJKV98}.
\item no manipulation of data must be applied and we need not refer to
 global or even local duality. That contributions of the type of power
 corrections, as suggested by the OPE, are negligible has been known for
 a long time.  This, however, does not proof the absence of other kind
 of non-perturbative effects. Therefore our conservative choice of the
 minimum Euclidean energy seems to be necessary.
\item
 as we shall see our non--perturbative ``remainder'' $\dah0$ is mainly
 sensitive to low energy data, which changes the chances of
 possible future experimental improvement dramatically, as illustrated 
in Fig.~\ref{fig:alpsta} (p-QCD errors not displayed).
\end{itemize}

\vspace*{48mm}

\section{The leading hadronic contribution to $\amu$.}  

\begin{figure}[h] \hspace*{5mm}
\begin{picture}(120,60)(-48,0)
{\scalebox{0.75 0.75}{%
\epsfbox{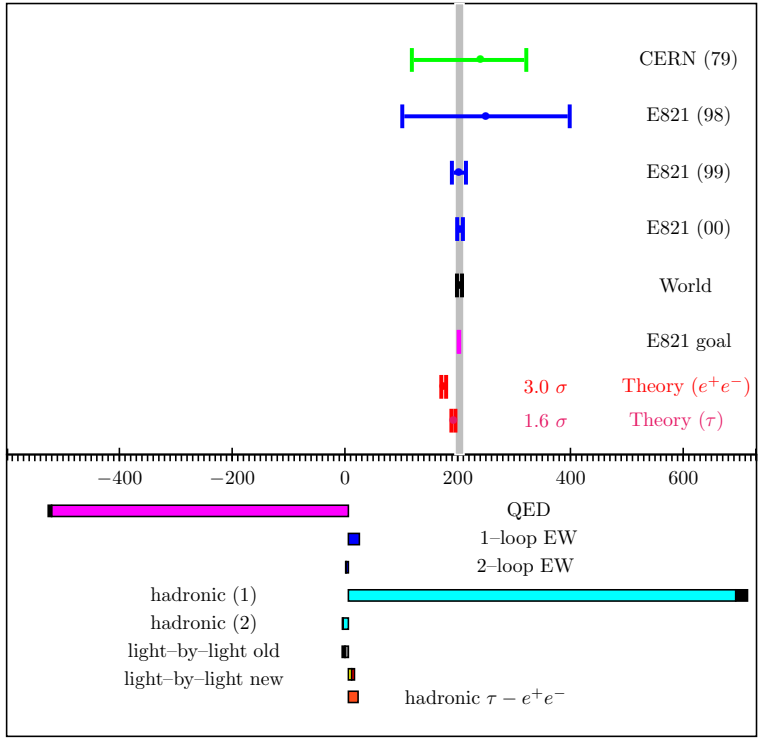}
}}
\end{picture} 
\vspace*{2mm}
\caption{\small  $a_\mu$-11659000 $\times 10^{-10}$: theory
vs. experiment in the year 2002 for (g-2) of the muon. The new E821
experiment at Brookhaven reviled a 2.7 $\sigma$ deviation from the
theory. The various types of SM contributions are shown in the lower
part of the figure.}
\label{fig:gm2status}
\end{figure}

The anomalous magnetic moment of the muon $\amu$ provides one of the
most precise tests of the quantum field theory structure of QED and
indirectly also of the electroweak SM. The precision measurement of
$a_\mu$ is a very specific test of the magnetic helicity flip
transition $\bar{\psi}_L\:{\sigma_{\mu \nu}}\: { F^{\mu\nu}}\:\psi_R$,
a dimension 5 operator which is forbidden for any species of fermions
at the tree level of any renormalizable theory. In the SM it is thus a
finite prediction which can be tested unambiguously to the extend that
we are able to calculate it with the necessary accuracy. For the
perturbative part of the SM an impressive precision has been reached.
Excitingly the new experimental result from Brookhaven~\cite{BNL}
which reached a substantial improvement in precision shows a 3.0[1.6]
$\sigma$ deviation from the theoretical prediction: $\left| a_\mu^{\rm
exp}-a_\mu^{\rm the}\right|=339(111)[167(107)] \times 10^{-11}$,
depending on whether one trusts more in an $e^+e^-$--data[$\tau$--data]
based evaluation of the hadronic vacuum polarization
contribution~\cite{DEHZ}. The status is illustrated in
Fig.~\ref{fig:gm2status}.  We refer to Ref.~\cite{CM01} for a recent
review and possible implications.

\vspace*{72mm}

Again contributions from virtual creation and reabsorption of strongly
interacting particles cannot be computed with the help of p-QCD and
cause serious problems. Fortunately the major such contribution again
enters via the photon vacuum polarization which can be calculated
along the lines discussed for the effective charge. The contribution
is described by the diagram\\

\hspace*{4.2cm} \epsfbox{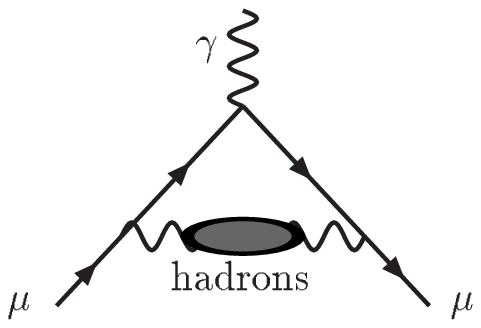}

\noindent
and is represented by the integral
\ba
\amuh = \left(\frac{\alpha m_\mu}{3\pi}
\right)^2 \bigg(\!\!\!\!\!\!\!\!\!
&&\int\limits_{4 m_\pi^2}^{E^2_{\rm cut}}ds\,
\frac{R^{\mathrm{data}}_\gamma(s)\;\hat{K}(s)}{s^2} \crn
\!\!\!\!\!\!\!\!\! &+& \int\limits_{E^2_{\rm cut}}^{\infty}ds\,
\frac{R^{\mathrm{pQCD}}_\gamma(s)\;\hat{K}(s)}{s^2}\,\, \bigg)
\label{AM}
\ea
which is similar to the integral (\ref{RS}), however with a different kernel
$K(s)$ which may conveniently be written in terms of the variable
\bea
x=\frac{1-\beta_\mu}{1+\beta_\mu}\;,\;\;\beta_\mu=\sqrt{1-4m^2_\mu/s}
\eea
and is given by
\ba
K(s)=\frac{x^2}{2}\:(2-x^2) 
    +\frac{(1+x^2)(1+x)^2}{x^2}
       \left(\ln(1+x)-x+\frac{x^2}{2} \right) 
    +\frac{(1+x)}{(1-x)}\:x^2 \ln(x)\;\;.
\label{KS}
\ea The integral (\ref{AM}) is written in terms of the rescaled
function \bea \hat{K}(s)=\frac{3 s}{m^2_\mu}K(s) \eea which is
bounded: it increases monotonically from 0.63 at threshold
$s=4m^2_\pi$ to 1 at $\infty$. Note the extra $1/s$--enhancement of
contributions from low energies in $\amu$ as compared to $\Delta
\alpha$.

\begin{figure}[h]
\begin{picture}(120,60)(-50,0)
{\scalebox{0.75 0.75}{%
\epsfbox{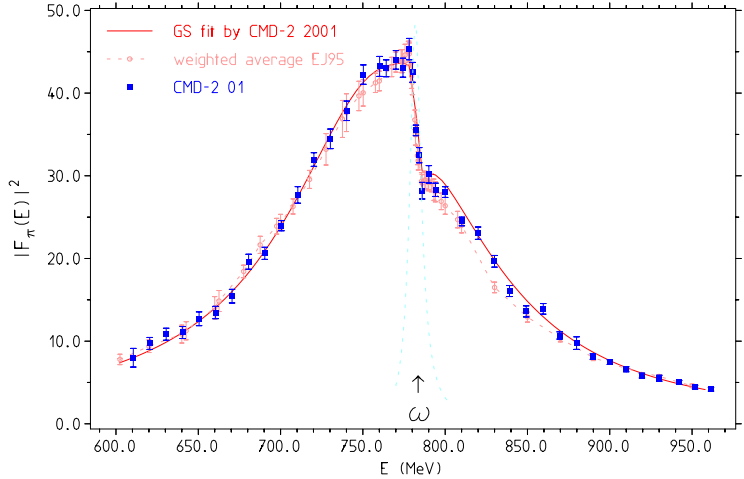}}}
\end{picture}

\caption{\small 
The dominating low energy tail is given by the channel $\epm \ra
\pi^+\pi^-$ which forms the $\rho$--resonance. We show a compilation
of the measurements of the square of the pion form factor
$|F_\pi(s)|^2=4\:R_{\pi\pi}(s)/\beta_\pi^3$ with
$\beta_\pi=(1-4m_\pi^2/s)^{1/2}$.
}
\label{fig:epemdata}
\end{figure}

\begin{figure}[h]
\begin{picture}(120,60)(40,15)
\epsfbox{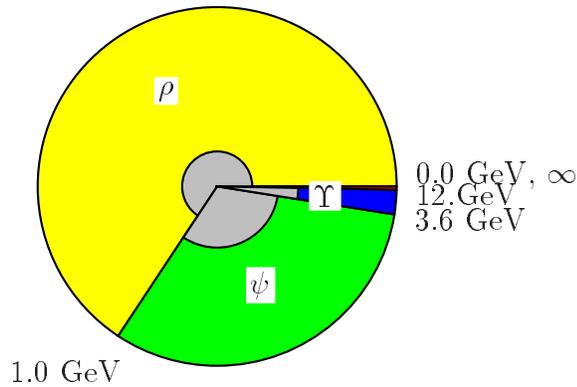}
\end{picture}

\caption{The distribution of contributions and errors
(shaded areas scaled up by 10) for $\amuh\;$. }
\label{fig:gmusta}
\end{figure}

A compilation of the $\epm$--data in the most important low energy
region is shown in Fig.~\ref{fig:epemdata}. The relative importance of
various regions is illustrated in Fig.~\ref{fig:gmusta}. The update of
the results~\cite{EJ95}, including the recent data from CMD and BES
yields \ba \amuh &=& (683.62 \pm 8.61) \times 10^{-10}\;\;. \label{eq:amuhad} \ea 
In Tab.~\ref{tab:amuerr} we summarize a few of the recent evaluations
of the leading hadronic contributions.
The most recent BNL $(g_\mu-2)$
measurement~\cite{BNL} gives \bea a_\mu^{\rm exp} &=& {(11659203~~ \pm
~8~~)} \times 10^{-10}~~~~~~~~ \mbox{ (world average)} \eea which
compares with the theoretical prediction\footnote{Recent new results concern the 
hadronic light-by-light contribution~\cite{Andreas} and the $O(\alpha^4)$ QED 
contribution to $a_e$~\cite{Kino02}} \bea a_\mu^{\rm the} &=& {(11659169.6 \pm
~9.4)} \times 10^{-10}~~~~~~~~~~~~~~ \mbox{ (SM)}\;. ~~~~~~~ \eea


\vspace*{46mm}

\vspace*{32mm}

\begin{table}[thb]
\begin{center}
\begin{tabular}{cl|cc}
\hline
$\amuh$ &$\damu$   & Method & Ref.\\
\hline
$6967 $&$156 $& data $< ~12.~~\gv$ + pQCD                   &\cite{EJ95,ADH98}\\
$6836 $&$~86 $& $\epm$ data incl. CMD-2 01, BES-II 99        &\cite{FJ02}      \\
$6821 $&$~86 $& Euclidean $> ~2.5~~\gv$        &\cite{FJ02}      \\
$7011 $&$~94 $& $\epm\:+\; \tau$  data $< ~12.~~\gv$ + pQCD &\cite{ADH98}     \\
$6951 $&$~75 $& data $< ~1.8~\gv$ and $3.3 - 5.0~\gv$ + pQCD&\cite{DH98a}     \\
$6924 $&$~62 $& in addition OPE (sum rules)                 &\cite{DH98b}     \\
$6847 $&$~70 $& $\epm$--based                 &\cite{DEHZ}     \\
$7019 $&$~61 $& $\tau$--based                 &\cite{DEHZ}     \\
$6831 $&$~62 $& $\epm$ data, pQCD driven, ``inclusive''    &\cite{HMNT}     \\
$ -   $&$~80 $& BNL $\amu$--experiment 2002                 &\cite{BNL}       \\
$ -   $&$~40 $& BNL final goal & \\
\hline
\end{tabular}
\caption{%
$\amuh$ and uncertainties in units $10^{-11}$.}
\label{tab:amuerr}
\end{center}
\end{table}
Let me comment on Tab.~\ref{tab:amuerr}: the first result with $\damu
\sim 156 \times 10^{-11}$ is based on $\epm$--data as analyzed in
Ref.~\cite{EJ95} and confirmed in~\cite{ADH98}. Perturbative QCD is
utilized only conservatively between 5.5 and 9.6 GeV and above 11 GeV
(see~\cite{ChHK00} for a discussion of the range of applicability of
p-QCD). Including the $\tau$--data from ALEPH Ref.~\cite{ADH98} finds
a result with 40\% improved uncertainty $\damu \sim 94 \times
10^{-11}$ under the assumption that iso-spin breaking is
negligible. Assuming the validity of p-QCD in the extended range
between 1.8 and 3.5 GeV and above 5 GeV ~\cite{DH98a} reduce the error
further to $\damu \sim 75 \times 10^{-11}$. In view of the bad quality
of the data in some ranges the idea to replace them by a theoretical
prediction is certainly able to lead to an improvement of the
evaluation. I do not see however, how to estimate reliably a
theoretical uncertainty in this approach since there are
non--perturbative effects around and the assumption of local duality,
i.e., $\sigma (\epm \ra {\rm hadrons}) \simeq \sigma (\epm \ra {\rm
quarks})$ in some average sense has no a priori theoretical
justification. Applying in addition the operator product expansion
(OPE) and sum rules to fix the quark and gluon condensate parameters
from the $\epm$--data in ~\cite{DH98b} a further reduction of the
error to $\damu \sim 62 \times 10^{-11}$ was claimed. That this
``improvement'' is obsolete has been clearly shown by the analysis
~\cite{EJKV98}. Once the full massive three loop
prediction~\cite{mqcd3} for the Adler function is compared with the
data it is impossible to establish any condensate effects. In the
region below 2.5 GeV where their contribution gets numerically
significant the perturbative expansion clearly is not reliable
anymore. The new analysis~\cite{DEHZ} is ``data--driven''
like~\cite{EJ95,ADH98} and confirms discrepancies between
$\epm$-- and $\tau$--data. The $\tau$--based result agrees with the
corresponding result of~\cite{ADH98}. If one would apply the
``theory--driven'' method of~\cite{DH98b} in conjunction with the new
CMD-2 data one would find a 4.5 $\sigma$ deviation from the
theoretical prediction: $\left| a_\mu^{\rm exp}-a_\mu^{\rm
the}\right|=426(95) \times 10^{-11}$ for the $\epm$--based approach.


A substantial improvement of the evaluation of $\amuh$ would be possible,
by including the $\tau$--data, provided one would understand iso--spin
violating effects sufficiently well~\cite{CEN}. This has been pioneered by
Ref.~\cite{ADH98}. Here one utilizes the fact that the vector--current
hadronic $\tau$--decay spectral functions are related to the
iso--vector part of the $\epm$--annihilation cross--section via an
iso-spin rotation: $$ \tau^- \ra X^-
\nu_\tau\;\;\; \leftrightarrow \;\;\; e^+ e¯ \ra X^0$$ where $X^-$ and
$X^0$ are related hadronic states. The $\epm$ cross--section is then
given by
\bea
\sigma_{\epm \ra X^0}^{I=1}= \frac{4 \pi \al^2}{s}v_{1,X^-}\;\;,\;\;\;
\sqrt{s} \leq M_\tau\;\;.
\eea

\vspace*{46mm}

\begin{figure}[th]
\begin{picture}(120,60)(-50,0)
{\scalebox{0.75 0.75}{%
\epsfbox{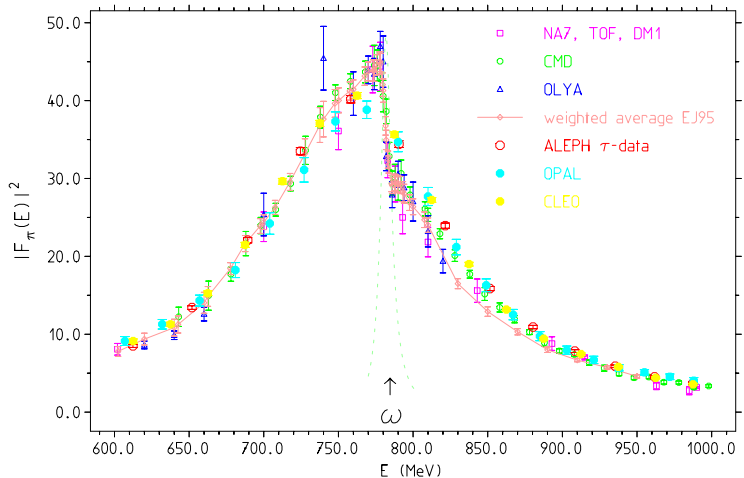}}}
\end{picture}

\caption{\small The square of the pion form factor $|F_\pi|^2$: a
compilation of the $\tau$--decay data in comparison with the
$e^+e^-$--annihilation data. The experimental cross sections are used
to calculate the leading hadronic contribution to (g-2) of the
muon.}
\label{fig:taudata}
\end{figure}

The $\tau$ spectral function $v_1$ is obtained from the normalized invariant
mass-squared distribution $\rho\:(X^-)\equiv (1/N_{X^-}\,dN_{X^-}/ds)$
of the $\tau$ vector channel $X^-\nu_\tau$ by
\bea
v_{1,X^-}=A\,
\frac{B(\tau^- \ra X^-\nu_\tau)}{B(\tau^- \ra e^-\nu_\tau\bar{\nu}_e)}\,
\rho\:(X^-)\,\bigg[\left(1-\frac{s}{M^2_\tau} \right)^2
\left( 1+\frac{2s}{M^2_\tau}\right) \bigg]^{-1}
\eea
where
\bea
A=\frac{M^2_\tau}{6|V_{ud}|^2\,(1+\delta_{\rm EW})}
\eea
with $|V_{ud}|=0.9752 \pm 0.0007$ the CKM mixing matrix element and
$\delta=0.0194$ the electroweak radiative corrections.  With the
precision of the validity of CVC, this allows to improve the $I=1$
part of the $\epm$ cross--section which by itself is not a directly
measurable quantity. It mainly improves the knowledge of the $\pipi$
channel ($\rho$--resonance contribution) which is dominating in
$\amuh$ (72\%). Fig.~\ref{fig:taudata} shows a compilation of the
$\tau$--data from ALEPH~\cite{ALEPH}, OPAL~\cite{OPAL} and
CLEO~\cite{CLEO} in comparison to the $\epm$--data. After taking into
account the known iso-spin breaking effects~\cite{CEN} the
$\tau$--data show substantial discrepancies in comparison with the
$\epm$--data (about 10\% just above the $\rho$--resonance). This issue
can certainly be settled by the radiative return experiments with
KLOE~\cite{KLOE} at LNF/Frascati and with
BABAR~\cite{Solodov:2002xu} at SLAC.

\section{Concluding Remarks}

Experimental efforts to measure very precisely the total cross section
{$\sigma(\epm \ra hadrons)$} at low energies are mandatory for the
future of electroweak precision physics. Taking into account recent
theoretical progress, these ``low energy'' measurements are not only
important for testing $\amu$ but as well for the effective fine
structure constant $\az$. A real breakthrough would be possible by
measuring $\sigma(\epm \ra hadrons)$ at 1\% accuracy below the
$\tau$--threshold.

Fortunately there is work in progress which can help to further reduce
the uncertainties of theoretical predictions: (i) CMD-2 and SND at 
VEPP-2M/Novosibirsk: can further improve up to 1.4 GeV. (ii)
KLOE at DA$\Phi$NE/Frascati: soon we expect a
measurement below the $\phi$ resonance which is 
competitive to the CMD-2 data~\cite{KLOE}.
(iii) BES at BEPC/Beijing: can
improve a lot in the important $J/\Psi$ region. 
(iv) Radiative return experiments at the $B$--factory
at SLAC with BABAR~\cite{Solodov:2002xu} can help a lot to improve the
problematic region between 1.4 and 2.0 GeV. (v) In future a
``$\tau$--charm facility'' tunable between 1.4 and 3.6 GeV would
settle the remaining problems essentially.

\section*{Addendum}
\newcommand{\dalh}{\Delta \alpha^{\rm had}}

In Ref.~\cite{EJKV98} it has been shown how one can obtain a better
control on the validity of pQCD by utilizing analyticity and looking
at to problem in the $t$--channel (Euclidean field theory approach). It
has been found that ``data'' may be safely replaced by pQCD at
$\sqrt{-t} \geq
2.5 \gv$. An application to the calculation of the running fine
structure constant has been discussed in~\cite{FJ98}. Here we consider
the application to the calculation of $\amuh$. Starting point is the
basic integral representation 
\be
\amuh=\frac{\alpha}{\pi}\int\limits_0^\infty\frac{ds}{s}
\int\limits_0^1 dx\:\frac{x^2\:(1-x)}{x^2+(1-x)\:s/m^2_\mu}\:
\frac{\alpha}{3\pi} \: R(s)\;\;.  
\ee 
If we first integrate over x we find the well known standard
representation as an integral along the cut of the vacuum polarization
amplitude in the time--like region, while an interchange of the order
of integrations yields an integral over the hadronic shift of the fine
structure constant in the space--like domain~\cite{LPdR72}:
\be
\amuh=\frac{\alpha}{\pi}\int\limits_0^1 dx\:(1-x)\: \dalh
\left(-Q^2(x)\right)
\label{RAI}
\ee
where $Q^2(x)\equiv \frac{x^2}{1-x}m_\mu^2$ is the space--like square
momentum--transfer or
$$x=\frac{Q^2}{2m_\mu^2}\:\left(\sqrt{1+\frac{4m_\mu^2}{Q^2}}-1\right)\;\;.$$
In this approach we (i) calculate the Adler function from the
$\epm$--data and pQCD for the tail above 13 GeV, (ii) calculate the
shift $\dalh$ in the Euclidean region with or without an additional
cut in the $t$--channel at 2.5 GeV and (iii) calculate $\amuh$
via~(\ref{RAI}).\\

Alternatively, by 
performing a partial integration in (\ref{RAI}) one finds
\be
\amuh=\frac{\alpha}{\pi}m_\mu^2 \int\limits_0^1 dx\:x\:(2-x)\:
\left(D(Q^2(x))/Q^2(x)\right)
\label{ADI}
\ee 
by means of which the number of integrations may be reduced by
one. The evaluation in both forms provides a good stability test of
the numerical integrations involved.

Utilizing the most recent compilation of the $\epm$--data we obtain
the result given in Tab.~\ref{tab:amuerr}. Not too surprisingly, as is
well known the contribution to $\amuh$ is dominated by the low energy
$\epm$--data below 1 GeV, here the replacement of data by pQCD does
not reduce the uncertainty. The reason is hat the pQCD contribution
replacing the Euclidean Adler function at $\sqrt{-t}> 2.5 \gv $ shows
a substantial uncertainty due to the uncertainty of the charm mass
$m_c(m_c)=1.15...1.35\;\gv$. The uncertainty in the strong coupling
constant $\alpha_s(M_Z^2)=0.120\pm0.003$ is small and is not the
dominating effect. In contrast to~\cite{DH98a} we do not obtain a
reduction of the error. Of course our cut at 2.5 GeV, which we think
is all we can justify, is more conservative than the 1.8 GeV in the
time--like region anticipated there. Thus the best value we can obtain
from presently available $\epm$--data alone is the result (\ref{eq:amuhad}).

\vspace*{6mm}

{\bf Acknowledgments\\~}

It is a pleasure to thank the organizers of the symposium {\it 50 Years of
Electroweak Physics} in honor of Professor Alberto Sirlin's 70th
birthday for the invitation and the kind hospitality.

\bb{99}

\bibitem{BNL} 
G.~W.~Bennett {\it et al}  [Muon g-2 Collaboration],
Phys.\ Rev.\ Lett.\  {\bf 89} (2002) 101804
[Erratum-ibid.\  {\bf 89} (2002) 129903].

\bibitem{LEP} The LEP Collaborations ALEPH, DELPHI, L3, OPAL {\it et al},
{\it Preprint} hep-ex/0101027 (unpublished).

\bibitem{Sirlin80} A. Sirlin, \PRD 22 (1980) 971.

\bibitem{TDR} TESLA, {\it Technical Design Report}, DESY-2001-000.

\bibitem{KalSab55}
G. K\"all\'en and A. Sabry, {\it K. Dan. Vidensk. Selsk. Mat.-Fys. Medd.}
{\bf 29} (1955) No. 17.

\bibitem{Ste98}
M. Steinhauser, {\it Phys. Lett.} B{\bf 429} (1998) 158.

\bibitem{GKL}
S.G. Gorishny, A.L. Kataev and S.A. Larin, {\it Phys. Lett.} B{\bf 259} (1991)
144;\\
L.R. Surguladze and M.A. Samuel, {\it Phys. Rev. Lett.} {\bf 66}
   (1991) 560; ibid. 2416 (Err),\\
K.G. Chetyrkin, {\it Phys. Lett.} {\bf B 391} (1997) 402.

\bibitem{ChK95}
 K.G. Chetyrkin and J.H. K\"uhn, \PLB {\bf 342} (1995) 356 and references therein.

\bibitem{ChHK00}
K.~G.~Chetyrkin, R.~V.~Harlander and J.~H.~K\"uhn,
Nucl.\ Phys.\ B {\bf 586} (2000) 56,
[Erratum-ibid.\ B {\bf 634} (2002) 413].

\bibitem{EJ95}
S.I.~Eidelman and F.~Jegerlehner, {\it Z. Phys.} {\bf C 67} (1995) 585;\\
F.~Jegerlehner, {\it Nucl. Phys. B} (Proc. Suppl.) {\bf 51C} (1996) 131.

\bibitem{CMD}
R.~R.~Akhmetshin {\it et al.}  [CMD-2 Collaboration],
Phys.\ Lett.\ B {\bf 527} (2002) 161.

\bibitem{BES}
J.~Z.~Bai {\it et al.}  [BES Collaboration],
Phys.\ Rev.\ Lett.\  {\bf 84} (2000) 594;
Phys.\ Rev.\ Lett.\  {\bf 88} (2002) 101802.

\bibitem{FJ01}
F.~Jegerlehner, ``The effective fine structure constant at TESLA
energies'', ECFA/DESY LC-TH-2001-035 Note, Februay 2001.

\bibitem{BP01} H.~Burkhardt and B.~Pietrzyk, Phys.\ Lett.\ B {\bf 513} (2001) 46.

\bibitem{DH98a} M. Davier, A. H\"ocker, \PLB {\bf 419}, 419 (1998)

\bibitem{KS98} J.H. K\"uhn, M. Steinhauser, \PLB {\bf 437}, 425 (1998)

\bibitem{GKNS98} S. Groote, J.G. K\"orner, N.F. Nasrallah, K. Schilcher,
                 \PLB {\bf 440}, 375 (1998)

\bibitem{DH98b} M. Davier, A. H\"ocker, \PLB {\bf 435}, 427 (1998)

\bibitem{Erler98} J. Erler, hep-ph/9803453

\bibitem{MOR00}
A.D. Martin, J. Outhwaite, M.G. Ryskin, \PLB {\bf 492} (2000) 69;
Eur.\ Phys.\ J.\ C {\bf 19} (2001) 681.

\bibitem{FJ86} F. Jegerlehner, \ZPC {\bf 32}, 195 (1986)

\bibitem{SVZ} M.A. Shifman, A.I. Vainshtein and V.I. Zakharov,
              \NPB {\bf 147}, 385 (1979)

\bibitem{mqcd3} K.G. Chetyrkin, J.H. K\"uhn, M. Steinhauser,
               \PLB {\bf 371}, 93 (1996);
               \NPB {\bf 482}, 213 (1996); B {\bf 505}, 40 (1997); 
               K.G.~Chetyrkin, R.~Harlander, J.H.~K\"uhn, M.~Steinhauser,
               \NPB {\bf 503}, 339 (1997)

\bibitem{JT98}
F.~Jegerlehner and O.V.~Tarasov, \NPB {\bf 549} (1999) 481.

\bibitem{FJ98} F. Jegerlehner, in ``Radiative Corrections'',
ed.~J. Sol\`a, World Scientific, Singapore, 1999

\bibitem{EJKV98} S.~Eidelman, F.~Jegerlehner, A.L.~Kataev, O.~Veretin,
\PLB {\bf 454} (1999) 369.

\bibitem{FJ02}
F.~Jegerlehner,
hep-ph/0104304v2 (updated);
talk at the workshop "Hadronic Contributions to the
Anomalous Magnetic Moment of the Muon", CPT, Marseille, March 2002.

\bibitem{DEHZ}
M.~Davier, S.~Eidelman, A.~H\"ocker and Z.~Zhang, hep-ph/0208177.

\bibitem{CM01}
A. Czarnecki, W. J. Marciano, hep-ph/0102122.

\bibitem{Andreas}
M.~Knecht and A.~Nyffeler,
Phys.\ Rev.\ D {\bf 65} (2002) 073034;
A.~Nyffeler,
hep-ph/0210347 (and references therein).

\bibitem{Kino02}
T.~Kinoshita and M.~Nio,
hep-ph/0210322.

\bibitem{ADH98} R. Alemany, M. Davier, A. H\"ocker,
                {\it Eur.~Phys.~J.} C {\bf 2}, 123 (1998)

\bibitem{HMNT}
K.~Hagiwara, A.~D.~Martin, D.~Nomura and T.~Teubner,
hep-ph/0209187.

\bibitem{ALEPH} R. Barate et al. (ALEPH Collaboration),
{\it Z. Phys.} C {\bf 76} (1997) 15; {\it Eur. Phys. J.} C {\bf 4} (1998) 409

\bibitem{OPAL} K. Ackerstaff et al. (OPAL Collaboration),
{\it Eur. Phys. J.} C {\bf 7} (1999) 571 

\bibitem{CLEO} S. Anderson et al. (CLEO Collaboration), 
{\it Phys. Rev.} D {\bf 61} (2000) 112002

\bibitem{CEN}
V.~Cirigliano, G.~Ecker and H.~Neufeld,
Phys.\ Lett.\ B {\bf 513} (2001) 361; JHEP {\bf 0208} (2002) 002.

\bibitem{KLOE}
A.~G.~Denig  [the KLOE Collaboration], hep-ex/0211024.

\bibitem{Solodov:2002xu}
E.~P.~Solodov  [BABAR collaboration], hep-ex/0107027.

\bibitem{LPdR72} B.E.~Lautrup, A. Peterman, E. de Rafael, {\it Phys. Rep.}{\bf 3}
(1972) 193.
 
\end{thebibliography}

\end{document}